\documentclass[printer]{aa}
\usepackage{graphics}
\usepackage{epsfig}

\begin{document}
%\thesaurus{07 (08.01.3; 08.06.3; 08.12.2)}

\title{The Liverpool-Edinburgh High Proper Motion Survey}
\author{Richard S. Pokorny\inst{1}, Hugh R. A. Jones\inst{1}, 
Nigel C. Hambly\inst{2}}
\offprints{rsp@astro.livjm.ac.uk}
\institute{Astrophysics Research Institute, Liverpool John Moores 
University, Twelve Quays House, Egerton Wharf, Birkenhead CH41 1LD, UK.\\
\and
Wide Field Astronomy Unit, Institute for Astronomy, 
University of Edinburgh, Blackford Hill, Edinburgh, EH9 3HJ, UK.}

\date{}
\authorrunning{R. S. ~Pokorny, H. R. A. ~Jones \& N. C. ~Hambly}

\abstract{We present a catalogue of 6206 stars which have proper 
motions exceeding 0.18$\arcsec$yr$^{-1}$ 
with an R-band faint magnitude limit of 19$\cdot$5mag.  
This catalogue has been produced using SuperCOSMOS 
digitized R-Band ESO and UK Schmidt Plates in 131 Schmidt 
fields covering more than 3,000 square degrees 
($>$7$\cdot$5$\%$ of the whole sky) at the South 
Galactic Cap.  The survey is $\ge$90$\%$ complete within the 
nominal limits of the Luyten Two Tenths Catalogue of 
$m_R$$\le$18$\cdot$5mag and 
0$\cdot$2$\le$$\mu$$\le$2$\cdot$5$\arcsec$yr$^{-1}$, 
and is $\ge$80$\%$ complete for $m_R$$\le$19$\cdot$5mag 
and $\mu\le$2$\cdot$5$\arcsec$yr$^{-1}$.
\keywords{Galaxy: kinematics and dynamics -- Galaxy: halo -- solar neighbourhood -- Stars: late-type -- 
subdwarfs -- white dwarfs} }
\maketitle

\section{Introduction}
\subsection{Proper Motion Surveys}

Proper motion surveys can give us a good insight into stellar populations and dynamics in the Solar 
neighbourhood.  They identify nearby objects, allow analysis of their space motions and, with data 
in more than one passband, give population differentiation.  These three factors mean that proper 
motion surveys are ideal in addressing the problems of the local stellar mass function and the halo 
mass function, both of which are currently poorly constrained (e.g. Henry et al., 1997, Lee 1993).  The 
problem with the local stellar mass function is in large part due to the undersampling of nearby 
stars upon which it is mainly based.  Identification of objects such as the nearby brown dwarf Kelu-1 
(Ruiz et al. 1997), which was discovered as part of the Calan-ESO proper motion survey (Ruiz et al. 
2001) and has an astrometrically estimated distance of $\sim$10pc, are vital in this respect.
Both present day and initial halo mass functions are poorly constrained (Lee 1993) due to a number 
of factors.  These include the relatively low number of nearby halo subdwarf 
stars and the difficulty in obtaining accurate distances for them.  However,  halo stars have 
greater space velocity and velocity dispersion than disc stars which means that a halo star will 
have a higher proper motion than a disc star at the same distance.  When combined with their 
subluminosity this provides an excellent method of differentiating them from disc stars (seeSect. 
5$\cdot$1).

However, all of the existing surveys suffer from one or more of the following problems;
\begin{itemize}
\item internal inhomogeneities which make statistical analysis difficult and unreliable,
\item low areal coverage which makes a survey subject to low number statistics,
\item use of non-contiguous survey fields which prevents cross-checking of discovered objects and 
completeness levels from areas where contiguous fields overlap.
\end{itemize}

One of the main purposes of this survey is to address these problems by conducting an internally 
homogeneous large area survey in contiguous fields by means of scripted computer algorithms.  By 
using only the R-band plates we also ensured that the survey would have a high completeness relative 
to other surveys which require an object to be present in more than one passband, thus missing faint 
objects with extreme colours, such as brown dwarfs and ancient halo subdwarfs and white dwarfs.  This 
is important because these objects are among the most astrophysically interesting to be selected by 
proper motion surveys.  Using only the R-band plates also means that our completeness is easier to assess.

\subsection{The Luyten Catalogues}

The largest catalogue of high proper motion stars to date is the New Luyten Catalogue of Stars with 
Proper Motions Greater Than Two Tenths of an Arcsecond (NLTT) (Luyten, 1979).  The NLTT and the Luyten 
Half Second Proper Motion Catalogue (LHS) (Luyten, 1979) were compiled over a period of $\sim$60 years 
from two main surveys; the Bruce Proper Motion Survey (BPM), covering an area from the South 
Celestial Pole to a declination of $\approx$+25$^{\circ}$, and the Palomar Observatory Sky Surveys 
(POSS I and II) covering the area from the North Celestial Pole to $\approx$--33$^{\circ}$.  The nominal 
faint limit of the Bruce survey is R$\simeq$16$\cdot$5mag and the faint limit of the POSS is 
R$\simeq$20mag (Morgan et al., 1992), the Luyten Catalogues have no stated bright limit.  The Luyten 
proper motion measurements were made using two different techniques; (i) manual plate blinking and (ii)
automatic machine detection (Williams, 2000).  The Luyten catalogues therefore contain some 
internal inhomogeneities such as a brighter limiting magnitude for fields south of --33$^{\circ}$ 
and variations in the completeness of the catalogues between different areas of the surveys.   
The stated proper motion limits of the NLTT are
0$\cdot$18$\arcsec$yr$^{-1}$$\le\mu\le$2$\cdot$5$\arcsec$yr$^{-1}$ whilst the LHS 
comprises stars from the NLTT with $\mu\ge$0$\cdot$5$\arcsec$yr$^{-1}$.

\subsection{Other Surveys and the Completeness of the Luyten Catalogues}

There have been a number of estimates of the completeness of the NLTT and LHS catalogues in 
recent years, mostly based on the results of small area proper motion surveys.

Scholz et al. (2000) conducted a survey in 40 fields using the Automatic Plate Measuring 
microdensitometer (APM) at Cambridge, UK.  They discuss the completeness of previous catalogues, as well 
as how they find far more stars in the relatively incomplete southern area of the NLTT, but beyond 
stating that the NLTT is incomplete they do not quantify their own completeness or, using their data, 
the completeness of the NLTT.

Evans (1992) compared the results from four fields of the APM proper motion project to the results 
for the same areas from the NLTT.  He discovered a number of NLTT stars which, when measured with 
the APM microdensitometer, proved to have proper motions below the 0$\cdot$18$\arcsec$yr$^{-1}$ 
cutoff limit.  He also found a larger number of stars which had proper motions above this 
limit which were not listed in the NLTT.  By conducting a Monte Carlo simulation he showed that 
these discrepancies could be almost entirely attributed to random measuring errors.  He further 
stated that these errors led to a contamination of between 10$\%$ and 20$\%$ and an incompleteness 
of $\sim$16$\%$ (ie a completeness of $\sim$84$\%$).  However, whilst the comparison was meticulously 
carried out and  extremely detailed it relied on comparison of a very small number of fields.  It 
is therefore possible that it suffers from selection effects and small number statistics.

Ruiz et al. (2001) compared their results from the Calan-ESO proper motion catalogue to the Luyten 
catalogues and concluded that for $\mu$$\leq$0.8$\arcsec$yr$^{-1}$ and R$\geq$13mag the LHS 
catalogue is $\sim$60$\%$ complete.  They also state that the Luyten Two Tenths catalogue (LTT) is 
$\sim$40$\%$ complete for R$\geq$13mag, but this refers to the 1975 version of the Two Tenths 
catalogue, which was superseded by the more complete NLTT in 1979.  The Ruiz et al. comparison 
suffers from a similar problem to that of Evans in that it used the relatively small number of 14 
fields.  In addition to this their survey was conducted by manual blink comparison which is always 
prone to human error, as highlighted by the larger incompleteness of the southern BPM section of 
the NLTT over the northern POSS section.  It was also conducted at low Galactic latitude 
$\mid$b$\mid$$\le$40$^{\circ}$ because fields at high Galactic latitude do not contain enough 
background stars for efficient blink comparison, which could be an important contributory factor in 
the incompleteness of the BPM section of the NLTT.

Dawson (1986) applied the V/V$_{\rm{max}}$ method as described by Schmidt (1968, 1975) to estimate a 
completeness of $\sim$90$\%$ for $\delta$$\ge$-33$^{\circ}$, $\mid$b$\mid$$>$10$^{\circ}$, 
R$\le$18mag and $\mu$$\ge$0$\cdot$5$\arcsec$yr$^{-1}$ (i.e. the northern areas of the LHS).

Flynn et al. (2001) applied a statistical test to the NLTT consisting of taking two concentric 
spheres centred around the Sun whose volumes are in the ratio 2:1 (thereby giving them a radial ratio 
of 1$\cdot$259 which corresponds to an average magnitude difference of 0$\cdot$5mag between the 
shells) and assuming that the completeness for stars in the inner sphere is 100$\%$.  It is then 
possible to calculate the completeness of the stars in the outer `shell', and by applying this to 
different magnitude bins they derive a plot of completeness for the LHS and NLTT catalogues relative 
to the completeness at R=13 (assumed for the purposes of the plot to be 100$\%$ complete).  This 
analysis shows the completeness of the NLTT dropping steadily from 80$\%$ at R=14 to 60$\%$ at 
R=18$\cdot$5mag.  However, Monet et al. (2000) suggest that this analysis is flawed since it 
does not take into account the different space densities of stars at high and low Galactic Latitude.  
Their recomputation of the Flynn et al. test gives a completeness at low Galactic latitude 
(15$\le$$\mid$b$\mid$$\le$35) of between 75$\%$ and 85$\%$ for 14$\le$R$\le$18mag.

Monet et al. (2000) conducted a survey in 35 fields using POSS II plates and estimated that the LHS is 
$\sim$90$\%$ complete, as well as recalculating the result of Flynn et al. (discussed above).

\section{The Data and Search Process}

\subsection{The SuperCOSMOS Sky Survey}

The SuperCOSMOS Sky Survey (SSS) (Hambly et al. 2001a) is a project of the Wide-Field Astronomy Unit 
at the Royal Observatory Edinburgh.  The purpose of the project is to digitise Schmidt survey plates 
from the ESO, UK and Palomar (southern area) Schmidt surveys with the SuperCOSMOS microdensitometer 
machine and make the digitised data available online to the general astronomical community.

\subsection{The Survey Data}

The proper motion survey was conducted by comparison of two epochs of SuperCOSMOS digitized 
R-band Schmidt plates.  The plates used are the ESO Schmidt and the Anglo-Australian UK Schmidt 
R-band surveys.  The two surveys were taken at different observatories by different groups which 
means that the epoch differences between the plates vary.  In the majority of cases the ESO 
plate was taken before the UK plate, although in a very small number of cases the UK plate is 
earlier.  The average epoch difference is 8$\cdot$5 years and the vast majority of fields have epoch 
differences between 5 and 13 years (see Fig. 2).  The survey was conducted using only the two 
R-band plates in order not to miss faint objects with extreme colours, such as late M dwarfs or brown 
dwarfs, which might not be found by a search which also required detections in the B$\rm{_J}$ and 
I band plates.  Once the survey was completed the positions of the candidate high proper motion stars 
were extrapolated to the B$\rm{_J}$ and I band plates in order to extract BRI colour information 
where possible.

\begin{figure*}
\begin{center}
\begin{minipage}{6in}
   \epsfig{file=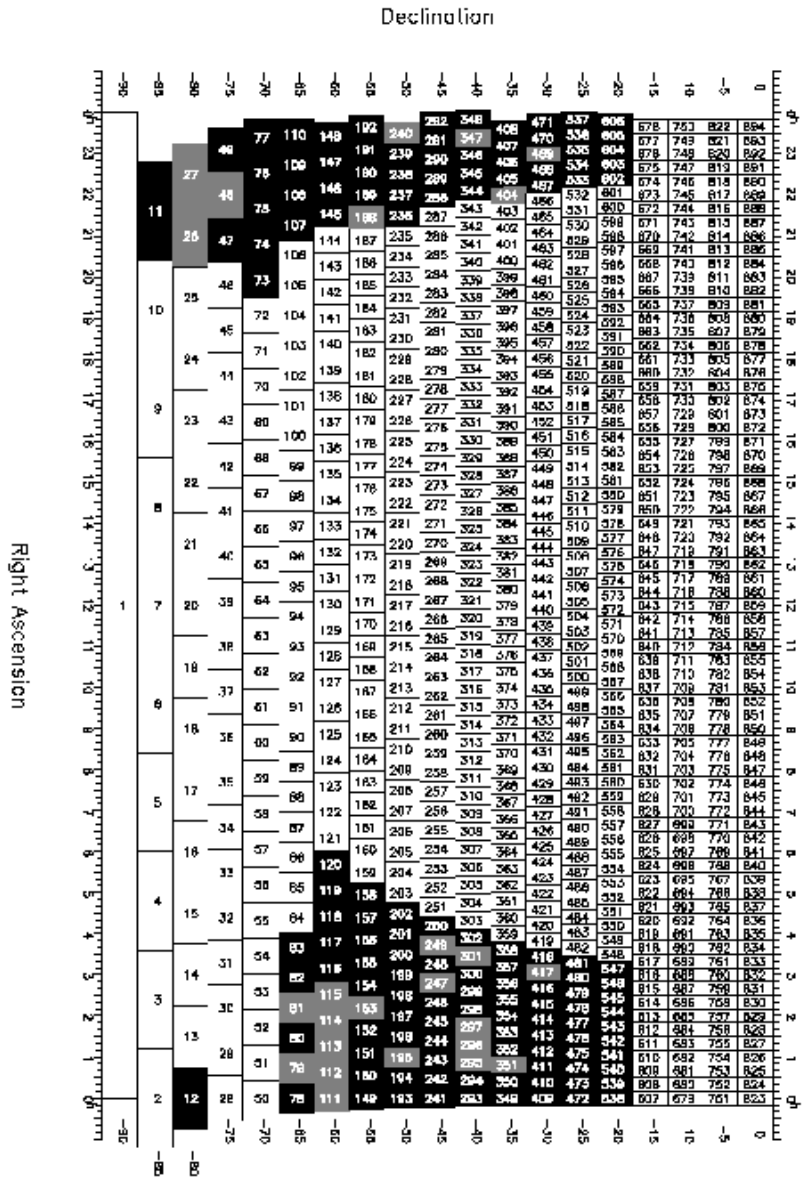,width=0.95\hsize,angle=0}
\end{minipage} 
  \caption{Plot of the fields used in this survey.  Fields shaded in dark were used, fields shaded in 
light were available, but could not be used either because of a short epoch difference between the ESO 
and UK plates or because they contained too few bright stars to calculate the full relative astrometry.}
\end{center}
\end{figure*}

\begin{figure}
\begin{center}
\begin{minipage}{6in}
   \epsfig{file=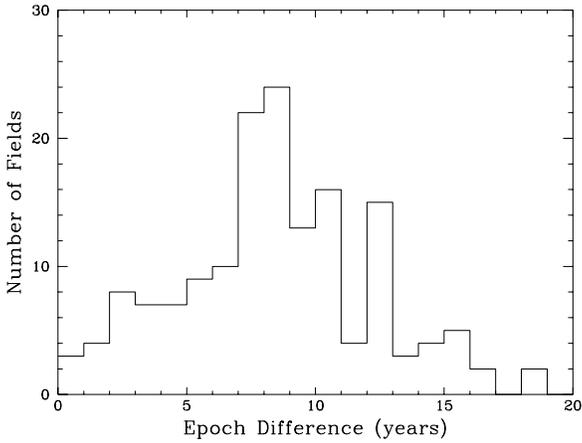,width=0.4\hsize,angle=270}
\end{minipage} 
  \caption{Histogram of the differences in epoch between the ESO and UK R-band plates for all of 
the 158 fields available to the survey.}
\end{center}
\end{figure}

\subsection{The Digitized Data}

The data has been digitized using the SuperCOSMOS microdensitometer at the Royal Observatory Edinburgh UK 
(Hambly et al. 1998) as part of the SuperCOSMOS Sky Survey (Hambly et al., 2001, a, b \& c).  
Each object on the plate is reduced to 32 image parameters which describe its position, size, intensity, 
etc., in a format known as Image Analysis Mode (IAM) (Hambly et al., 2001a).  These parameters make 
it possible to perform full astrometry and photometry on the data and to compare the data from two or 
more plates of the same field.

\subsection{The Search Programs}

The survey was conducted by a series of programs which manipulate the IAM files to perform full relative 
astrometry on the two epochs of each field.  The programs attempt to pair all objects between the two 
plates from which the high proper motion stars may be selected.  Objects are selected for the search 
process on the basis of a number of the IAM image parameters; 
\begin{itemize} 
\item the quality flag (which indicates if there are any problems with the object such as it being too 
faint, too bright, too large or too small), 
\item the deblending flag which indicates if the object is part of a larger object (e.g. part of the 
diffraction spike or halo of a bright star), 
\item the classification flag, which indicates whether the object is a galaxy, a star, dust or 
dirt, or too faint to classify.
\end{itemize}

The first task is to match the coordinate system of the UK plate to that of the ESO plate and trim the 
edges from the data (the latter being important because the UK plates are 6$^\circ$ square whilst the 
ESO plates are 5$^\circ$ square).  The data for both plates is then reordered within the IAM files into 
24 lanes so that the pairing programs only need to search a limited area of the plates rather than the 
whole of the dataset thus saving computer time and resources.  

The position of each object on the UK plate is then checked against the ESO plate for any object at the 
same position.  This is a simple first pass procedure to identify objects which have $\sim$ zero 
proper motion.  Every ESO plate object left unpaired by this procedure is searched for on the UK plate 
in annuli of gradually increasing radius.  The search looks for any object which has similar IAM image 
parameters to that of the object on the ESO plate.  When a match is found an identifier is written to a 
file of paired objects and the program moves on to the next object.  If no match is found within the 
cutoff radius the program writes a null identifier to the pairs file.  

The search is conducted in this two step manner because the likelihood of finding a match decreases as 
the search radius increases.  Searching the entire area out to the cut-off radius in one pass would use 
a large amount of computer time and processing power, which would be mostly wasted since for the vast 
majority of stars a match will be found within the first few annuli.  Additionally, searching the full 
radius in one go could result in two or three objects being identified as possible pairings for every 
ESO object, requiring the software to choose between these objects as to which is the most likely to be 
the correct one.  Again, this would again be a waste of time and processor power, since the most likely 
match of two or more similar candidates would always be the one nearest to the first epoch position.  

We then perform full relative astrometry on the stars in x and y plate coordinates (measured 
in mm).  The next stage is to calculate a transformation from x-y to RA-dec coordinates, which can then 
be applied to the x-y relative astrometry to yield full relative proper motions in RA and dec.  Once this 
has been done catalogue stars can be selected by applying proper motion cuts to the paired objects.  
Finally the catalogue can be compared to an electronic version of the NLTT (see Sect. 4).

Limits of 0$\cdot$18$\arcsec$yr$^{-1}$$\le\mu\le$20$\cdot$0$\arcsec$yr$^{-1}$, and 
9$\le$R$\le$19$\cdot$5mag were applied to the final selection of proper motion objects.  The lower 
proper motion limit was chosen to avoid any erroneous pairings due to the positional errors in the 
SuperCOSMOS data and to be consistent with the NLTT.  The high proper motion limit was chosen to be well 
above any expected proper motion in order to include all potential objects.  The bright magnitude limit was 
chosen because of the problems encountered with bright objects on a Schmidt plate (e.g. size of image, 
diffraction and saturation), whilst the faint magnitude limit was chosen because it becomes extremely 
difficult to differentiate by eye between dust and true stellar images for R$>$19$\cdot$5mag on a Schmidt 
plate, which must always be the final test for any proper motion object discovered from the digitized data.  
This has the additional effect that the object classification contained in the IAM data becomes less 
accurate for faint objects which also makes selection of good pairs by use of the programs extremely 
difficult for images fainter than R=19$\cdot$5mag.  

After the proper motion selection was completed the data was compared to the B$\rm{_J}$ and I band 
plates (where available) in order to obtain magnitude data for the objects in those passbands.

\section{The Catalogue}

The catalogue comprises 6605 objects, 399 of which are extra detections of stars from the plate overlap 
regions and are already listed in the catalogue.  This means that there are actually 6206 high proper 
motion stars in the catalogue.  They lie within an area of -82.5$^{\circ}$$\le$dec$\le$-22.5$^{\circ}$ 
and 21hrs$\le$RA$\le$5hrs with magnitude limits of 9$\leq$m$_{R}$$\leq$19.5mag and proper motion limits 
of 0.18$\leq$$\mu$$\leq$20$\arcsec$yr$^{-1}$.  
Table 1 is a sample from the catalogue, comprised of the first 40 objects. 

The columns are; catalogue number, right ascension (hr, m, s) and declination ($^{\circ}$, $\arcmin$, 
$\arcsec$) (equinox J2000 at the epoch of the ESO plate), multiple detection flag, ESO/SRC number of 
the field the object was located on, Epoch of ESO observation, proper motion in R.A., error on proper 
motion in R.A., proper motion in dec, error on proper motion in dec, total proper motion, error on 
total proper motion (all proper motions and their associated errors are given in $\arcsec$yr$^{-1}$), 
position angle of proper motion (in degrees), apparent magnitudes: B$\rm{_J}$, R$\rm{_{ESO}}$, R$\rm{_{UK}}$, I.  The 
multiple detection flag indicates where, in an overlap region between plates, an object is identified 
more than once (in some cases four times).  This flag is 0 if the object is not a multiple detection or 
has the more accurate data of the mutiple detections.  The flag is 1 if the object is a multiple 
detection with less accurate data, e.g. where there is no matching object was identified on the B$\rm{_J}$ or 
I band plates or the proper motion error is higher than for other detections.  In those cases where no 
match could be found in the B$\rm{_J}$ or I band plates the apparent magnitude is given as 0.00.

\begin{table*}
 \centering
 \begin{minipage}{140mm}
  \caption{Sample table from the catalogue.  Columns are; catalogue number, R.A., dec, multiple detection 
flag, SERC field number, ESO R-band plate epoch, $\mu$$_{\alpha}$, error on $\mu$$_{\alpha}$, $\mu$$_{\delta}$, 
error on $\mu$$_{\delta}$, total $\mu$, error on $\mu$, position angle, magnitudes - B$\rm{_J}$, R$\rm{_{ESO}}$, R$\rm{_{UK}}$, I.}
\rotatebox{270}{\begin{tabular} {|r|rrr|rrr|c|r|c|c|c|c|c|c|c|r|c|c|c|c|} \hline
N$_C$&\multicolumn{3}{c|}{R.A.}&\multicolumn{3}{c|}{dec}&F$_M$&Fld&Epoch&$\mu$$_{\alpha}$&$\sigma$$_{\mu}$$_{\alpha}$&
$\mu$$_{\delta}$&$\sigma$$_{\mu}$$_{\delta}$&$\mu$&$\sigma$$\mu$&P.A.&B$\rm{_J}$&R$\rm{_{ESO}}$&R$\rm{_{UK}}$&I\\ \hline
 1&0&0& 0.364&-46& 1& 4.89&0&292&1989.513& 0.1662&0.0184&-0.0928&0.0192&0.1894&0.0265&117.42&19.55&17.56&17.66&16.87\\
 2&0&0& 7.346&-40&39&52.36&0&348&1985.774&-0.0446&0.0182&-0.1995&0.0198&0.2035&0.0267&191.33&17.82&15.79&15.92&14.76\\
 3&0&0& 7.353&-40&39&52.90&1&293&1988.853&-0.0228&0.0412&-0.2003&0.0251&0.2007&0.0480&186.97&17.73&15.70&15.70&14.50\\
 4&0&0&12.691&-31&46&19.22&0&409&1981.750& 0.1856&0.0472&-0.0398&0.0532&0.1887&0.0707&102.41&14.23&12.05&12.25&11.31\\
 5&0&0&18.360&-27& 8&36.13&1&409&1981.750& 0.2766&0.0300&-0.0896&0.0359&0.2892&0.0465&108.19&14.72&12.32&12.65&11.42\\
 6&0&0&18.448&-27& 8&36.32&0&537&1986.806& 0.2626&0.0175&-0.0732&0.0096&0.2714&0.0199&104.72&14.70&12.36&12.65&11.32\\
 7&0&0&18.488&-27& 8&36.59&1&471&1989.530& 0.2417&0.0253&-0.1389&0.0958&0.2774&0.0986&118.54&14.72&12.41&12.36&11.33\\
 8&0&0&18.527&-27& 8&36.86&1&472&1990.728& 0.2654&0.0424&-0.0697&0.0440&0.2731&0.0608&104.93&14.55&12.20&12.23&10.76\\
 9&0&0&19.380&-26&12&48.55&1&537&1986.806& 0.2374&0.0180&-0.0420&0.0188&0.2399&0.0259& 99.22&19.75&17.35&17.52&15.69\\
10&0&0&19.455&-26&12&48.80&0&472&1990.728& 0.2140&0.0179&-0.0337&0.0183&0.2156&0.0255& 99.18&19.56&17.31&17.43&15.54\\
11&0&0&22.130&-41&33&45.67&1&293&1988.853&-0.0444&0.0395&-0.2279&0.0260&0.2310&0.0471&191.45&17.58&15.45&15.76&14.23\\
12&0&0&22.135&-41&33&45.01&0&348&1985.774&-0.0551&0.0236&-0.2289&0.0228&0.2343&0.0326&192.23&17.72&15.60&15.74&14.01\\
13&0&0&24.175&-39& 4& 4.21&0&348&1985.774& 0.1409&0.0373&-0.1173&0.0234&0.1825&0.0438&128.49&17.01&14.66&14.89&12.79\\
14&0&0&24.226&-39& 4& 4.64&1&293&1988.853& 0.1571&0.0546&-0.1160&0.0391&0.1944&0.0669&126.85&16.73&14.54&14.73&13.03\\
15&0&0&25.643&-18&46& 1.64&0&538&1984.788& 0.2006&0.0122& 0.0125&0.0199&0.2000&0.0232& 86.65&18.69&16.50&16.62&14.40\\
16&0&0&25.652&-18&46& 1.60&1&606&1985.626& 0.1974&0.0268& 0.0243&0.0324&0.1980&0.0418& 82.20&18.39&15.99&16.07&13.89\\
17&0&0&26.931&-41&25&30.40&0&348&1985.774& 0.4938&0.0236&-0.0489&0.0228&0.4939&0.0326& 94.29&17.70&15.74&15.65&14.57\\
18&0&0&27.089&-41&25&30.44&1&293&1988.853& 0.5038&0.0395&-0.0456&0.0260&0.5035&0.0471& 95.60&17.68&15.54&15.68&14.68\\
19&0&0&30.500&-26&13&54.54&1&537&1986.806& 0.2913&0.0557& 0.0861&0.0379&0.3023&0.0671& 72.68&14.53&12.24&12.63&11.12\\
20&0&0&30.594&-26&13&54.26&0&472&1990.728& 0.3026&0.0443& 0.0954&0.0322&0.3158&0.0545& 72.72&14.20&11.72&12.09&10.74\\
21&0&0&32.679&-53&26& 4.18&0&149&1989.597& 0.1754&0.0173&-0.1351&0.0167&0.2204&0.0239&127.94&18.92&16.61&16.75&14.53\\
22&0&0&33.945&-36&50& 8.79&0&349&1988.788& 0.4433&0.0469& 0.0866&0.0497&0.4495&0.0680& 79.27&16.14&14.06&14.21&12.84\\
23&0&0&35.458&-45&15& 5.96&0&292&1989.513& 0.1913&0.0481&-0.0289&0.0299&0.1925&0.0564& 96.72&14.28&12.69&12.84&12.16\\
24&0&0&36.181&-47& 7&50.26&0&292&1989.513& 0.1974&0.0203&-0.0920&0.0204&0.2167&0.0286&113.07&20.53&18.07&18.28&17.25\\
25&0&0&36.221&-47& 7&50.34&1&193&1989.583& 0.1795&0.0256&-0.0932&0.0224&0.2013&0.0339&117.79&20.72&18.30&18.47&17.41\\
26&0&0&36.735&-40&29&16.69&1&348&1985.774& 0.1975&0.0374&-0.0684&0.0365&0.2080&0.0520&107.76&15.70&14.39&14.45&13.90\\
27&0&0&36.799&-40&29&16.83&0&293&1988.853& 0.2033&0.0379&-0.0613&0.0240&0.2113&0.0446&107.17&15.69&14.44&14.44&13.79\\
28&0&0&37.350&-24&38&26.40&0&472&1990.728&-0.1293&0.0309&-0.2127&0.0440&0.2477&0.0535&211.49&16.95&14.57&14.67&12.69\\
29&0&0&37.401&-24&38&25.76&1&537&1986.806&-0.1489&0.0452&-0.1858&0.0345&0.2370&0.0566&217.88&17.12&14.92&15.18&13.18\\
30&0&0&45.060&-62&43&41.48&0& 78&1988.689& 0.2170&0.0509&-0.2317&0.0472&0.3159&0.0691&137.21&13.34&11.13&11.46&10.28\\
31&0&0&47.988&-31&31&33.46&0&409&1981.750& 0.0063&0.0641&-0.1844&0.0244&0.1835&0.0682&178.25&12.90&11.95&12.28&11.83\\
32&0&0&51.837&-33&53&14.74&0&349&1988.788& 0.1803&0.0286& 0.0156&0.0239&0.1801&0.0371& 85.31&18.59&16.45&16.53&14.55\\
33&0&0&55.111&-53&19&47.56&0&149&1989.597& 0.1781&0.0546&-0.0736&0.0171&0.1918&0.0570&112.71&17.12&14.81&14.94&13.30\\
34&0&0&56.293&-79&51&22.35&0& 12&1984.739& 0.2041&0.0132& 0.0857&0.0088&0.2203&0.0158& 67.62&17.81&15.61&15.76&13.67\\
35&0&0&59.144&-25&41&39.11&0&537&1986.806& 0.0082&0.0309&-0.2114&0.0188&0.2105&0.0360&176.91&16.19&14.64&14.74&13.92\\
36&0&1& 1.621&-21&48&51.16&0&538&1984.788& 0.0764&0.0263&-0.2299&0.0187&0.2411&0.0321&161.77&17.15&15.33&15.23&13.24\\
37&0&1& 4.854&-33&57&39.33&0&349&1988.788& 0.2701&0.0666&-0.0838&0.0509&0.2815&0.0835&107.46&16.75&14.60&14.56&13.05\\
38&0&1&26.461&-42&53& 8.67&1&292&1989.513& 0.1973&0.0265&-0.0825&0.0289&0.2128&0.0390&110.76&16.18&14.26&14.25&13.32\\
39&0&1&26.499&-42&53& 8.58&0&241&1989.806& 0.1836&0.0154&-0.0504&0.0223&0.1895&0.0270&105.53&16.27&14.31&14.49&13.55\\
40&0&1&27.468&-30&36&31.17&0&409&1981.750& 0.1509&0.0341&-0.1498&0.0178&0.2114&0.0383&134.92&16.77&14.88&14.78&14.04\\
\hline
\end{tabular}}
\end{minipage}
\end{table*}

\section{Completeness of the Survey}

\subsection{The completeness test}
Our completeness test consists of placing 100 simulated proper motion stars, each of which has 
the same magnitude and proper motion, into the data set for an ESO field and conducting the search 
process to see how many are recovered.  This is repeated for the same magnitude with increasing 
proper motion starting from 0$\cdot$3$\arcsec$yr$^{-1}$ then 0$\cdot$5$\arcsec$yr$^{-1}$ and then 
in steps of 0$\cdot$5$\arcsec$yr$^{-1}$ up to 10$\arcsec$yr$^{-1}$.  The same procedure is then 
repeated for every 0$\cdot$5 magnitudes from R=12mag to R=19mag.  

The lower limit of 0$\cdot$3$\arcsec$yr$^{-1}$ was used because the simulated stars are added to the 
data after the coordinate matching procedure but before the astrometric solution is applied to the data.  
The calculated proper motions of the simulated stars can therefore vary by more than 10$\%$, which 
means that an input proper motion of 0$\cdot$2$\arcsec$yr$^{-1}$ results in some of the 
simulated stars being lost due to their calculated proper motions falling below the lower proper 
motion cutoff limit of 0$\cdot$18$\arcsec$yr$^{-1}$.  The upper limit of 10$\arcsec$yr$^{-1}$ was 
chosen because no candidate star was discovered by the survey with a proper motion above 
$\sim$10$\arcsec$yr$^{-1}$.

One of the main problems with assessing the overall completeness of this survey is the way in which 
the plate epoch differences vary from field to field.  An increase in epoch difference means that, 
for a given proper motion, the physical image separation of a star between the two epochs will also 
increase.  This results in an increase in the probability of an erroneous pairing which will in turn 
reduce the completeness.  Fortunately this problem can be quantified by running the test on fields 
with different epoch differences.  For this reason the tests were carried out on three fields;
\begin{itemize}
\item Field 188, with a 2$\cdot$980 year epoch difference,
\item Field 479, with a 9$\cdot$078 year epoch difference,
\item Field 158, with a 16$\cdot$911 year epoch difference,
\end{itemize}
These fields are representative of the shortest and longest epoch differences as well as an 
epoch difference roughly halfway between these two extremes.

\subsection{Epoch difference limits on fields used in the survey}
Fields with epoch differences of less than $\sim$3 years were discarded due to a large increase 
in false pairings at these shorter epoch differences.  It was originally thought that these false 
pairings were caused by dirt and dust which are randomly distributed on the plates.  However, 
there are too few dust particles on the UK plates which could be paired with the stars on the ESO 
plates to have such a large effect on the number of pairings.  The overall proper motion errors 
are, on average, about 0.04$\arcsec$yr$^{-1}$, but analysis of these errors reveals that they 
increase for fields with a short epoch difference.  The proper motion error is propogated from the 
centroiding errors on the SuperCOSMOS data.  This increase begins to have a significant 
effect on the proper motions at an epoch difference of slightly less than 3 years.  The falsely 
identified stars are therefore not wrongly paired but are the result of the higher proper motion 
errors on stars with proper motions which are actually below the 0.18$\arcsec$yr$^{-1}$ cut off 
limit in plates with short epoch differences.

Although the completeness decreases for plates with large epoch differences (see Fig. 3) this 
is simply a systematic effect of the increase in the plate to plate separations of stars at any 
given proper motion increasing the likelihood of false pairings.  Since this only affects faint 
objects with relatively large proper motions no upper limit was applied to the allowed epoch 
difference of plate pairs.

\subsection{Completeness test result}
Fig. 3 shows that for fields with a short epoch difference the test completeness remains over 
90\% for all magnitudes and proper motions tested.  Fig. 3 shows how the completeness decreases 
for increasing epoch difference and Fig. 4 how it decreases for R$>$18.5mag.  This drop off is due 
partly to the increasing number of stars at faint magnitudes, and partly to dust and dirt on the 
plates, which can be difficult to distinguish from faint stars.  The former results in an increase 
in the density of faint stars which in turn increases the probability of a false pairing.  The latter 
is only a small effect for the reasons discussed above.  Morgan et al. (1992) give the limiting 
magnitudes of both ESO R and UK R plates as 22, although they point out that the limiting magnitudes 
of Schmidt plates are not defined in a consistent manner.  Hambly et al. (2001b) give the limit for 
identifying an object above the noise of an ESO Schmidt plate as R$\sim$21.  Our limit of R=19.5mag 
is therefore a little conservative, but appropriate since we need to do more than just identify 
objects above the noise and are aiming for a uniform catalogue.

\begin{figure}
\begin{center}
\begin{minipage}{6in}
   \epsfig{file=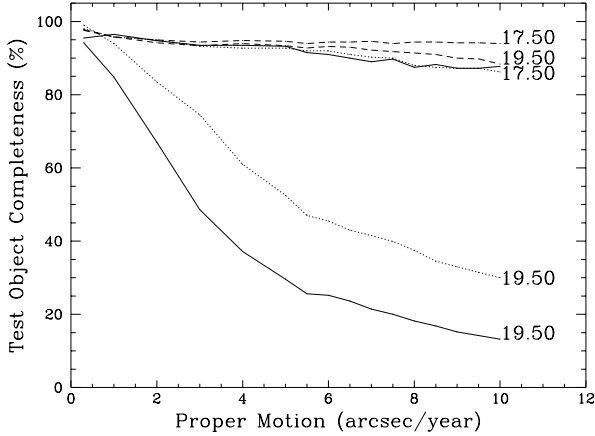,width=0.4\hsize,angle=270}
\end{minipage} 
  \caption{Percentage completeness of the Survey for fields with epoch differences of
3 years (field 188, dashed line), 9 years (field 479, dotted line) and 17 years (field 
158, solid line).  R band magnitudes are shown at the right hand ends of the lines.}
\end{center}
\end{figure}

\begin{figure}
\begin{center}
\begin{minipage}{6in}
   \epsfig{file=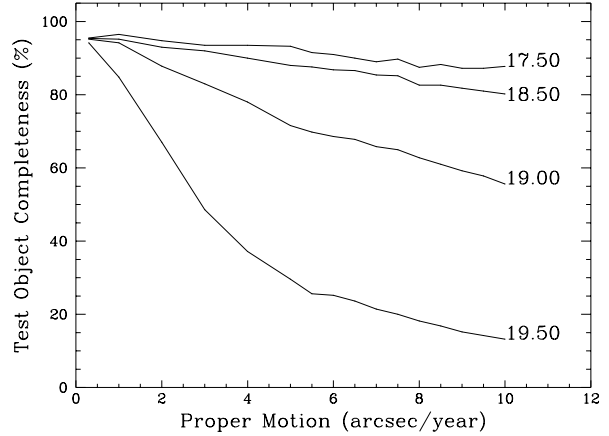,width=0.4\hsize,angle=270}
\end{minipage}
  \caption{Percentage completeness for field 158 with a 17 year epoch difference.  R 
band magnitudes are shown at the right hand end of each line.}
\end{center}
\end{figure}

\subsection{Cumulative number of proper motion stars}
Another measure of the completeness of the survey is relationship between the proper motion and 
the cumulative number of stars with proper motions equal to or greater than a given proper motion.  
For any tangential velocity, V$_T$,
\begin{equation}
\mu \propto r^{-1}  
\end{equation}
the volume (V) is proportional to r$^3$. \\
For a uniform distribution of stars the number of stars per unit volume is constant (i.e. $\rho$ 
is constant).  Since 
\begin{equation}
Nc=\rho V 
\end{equation}
this implies that 
\begin{equation}
Nc \propto r^3 
\end{equation}
and, therefore, that 
\begin{equation}
Nc \propto \mu ^{-3}.  
\end{equation}
Fig. 5 is plot of Nc against $\mu$ for all stars in this survey with $\mu$$\ge$0$\cdot$2$\arcsec$yr$^{-1}$.  
The data has a slope of -3.007 for 0$\cdot$2$\le$$\mu$$\le$1$\cdot$0$\arcsec$yr$^{-1}$ and demonstrates that 
this survey is complete within these limits.

\begin{figure}
\begin{center}
\begin{minipage}{6in}
   \epsfig{file=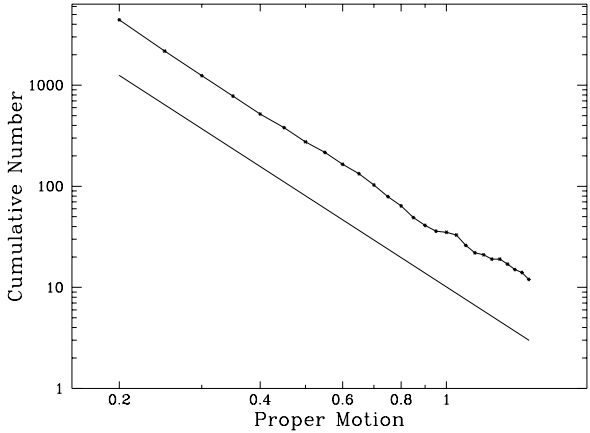,width=0.4\hsize,angle=270}
\end{minipage} 
  \caption{Cumulative number of stars in this survey as a function of their proper motions.  
The offset straight line has a slope of -3.}
\end{center}
\end{figure}

\subsection{Completeness of the NLTT}

This survey identified 6206 high proper motion objects in $\sim$7$\cdot$5$\%$ of the whole sky with a 
completeness of $\sim$90$\%$.  This implies $\sim$86,400 stars with magnitudes 9$\leq$R$\leq$19.5mag 
and proper motions $\mu$$\geq$0.18$\arcsec$yr$^{-1}$ over the whole sky compared to the 58,845 stars 
in the NLTT, which would suggest at first sight that the overall completeness of the NLTT is $\sim$70$\%$.  
This is, however, extremely misleading since the NLTT is made up of two surveys (see Sect. 1$\cdot$2).  
The POSS plate section of the survey (north of dec$\sim$-33$^{\circ}$) has a nominal faint magnitude limit 
of R$\sim$18$\cdot$5mag, and was mostly conducted by a laser microdensitometer machine, thus making it 
largely free from human error.  The BPM section (south of $\sim$-33$^{\circ}$) has a nominal faint R-band 
magnitude limit of R$\sim$16$\cdot$5mag and was conducted entirely by eye on a blink comparator (which 
Luyten himself developed).  The Luyten catalogues are therefore far less complete in the southern BPM area, 
as well as having a brighter limiting magnitude.  Figs. 6 and 7 show how the completeness and faint 
limits of the different sections of the NLTT vary.  Fig. 6 shows the number of high proper motion stars 
in the NLTT and this survey for the 43 fields in this survey which have field centres with 
dec$\geq$-~30$^{\circ}$.  Fig. 7 shows the number of high proper motion stars in the NLTT and this survey 
for the 73 fields in this survey which have field centres with dec$\leq$-40$^{\circ}$.  The overlap between 
fields with centres at dec=-35$^{\circ}$ and dec=-30$^{\circ}$ means that the fields with centres at 
dec=-35$^{\circ}$ contain a number of stars discovered from the POSS plates which would increase the number 
of stars in the NLTT relative to this catalogue.  This would give a false indication of the relative numbers 
of stars between the two catalogues, and for this reason the 15 fields with plate centres at dec=-35$^{\circ}$ 
were omitted from this plot.

As discussed in Sect. 1$\cdot$3, Monet et al. have also shown that the completeness of the NLTT is dependent on 
Galactic latitude.  Our survey was conducted at Galactic latitudes -30$^{\circ}$$\ge$b$\ge$-90$^{\circ}$ 
which most nearly corresponds to the high Galactic latitude fields which Monet et al. show to be less 
complete.  Our analysis shows that in the POSS section of the NLTT is $\sim$85$\%$ complete  for 
10$\cdot$5$\le$R$\le$19$\cdot$5mag and that the BPM section is $\sim$90$\%$ complete for R$\le$13mag with 
the completeness dropping off rapidly for fainter magnitudes.

\begin{figure}
\begin{center}
\begin{minipage}{6in}
   \epsfig{file=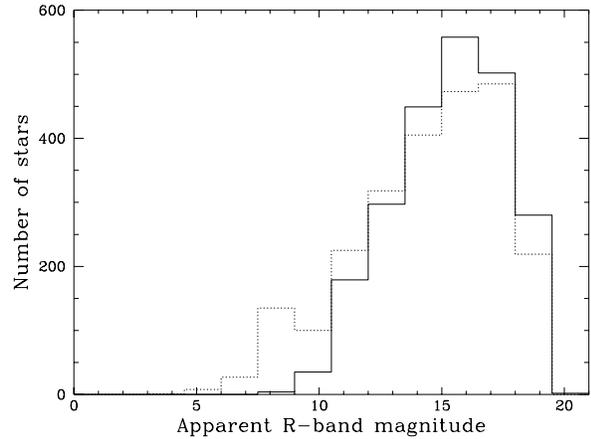,width=0.4\hsize,angle=270}
\end{minipage} 
  \caption{Histogram of the number of stars with $\mu$$\ge$0$\cdot$2$\arcsec$yr$^{-1}$ for 
the NLTT (dashed line) and this catalogue (solid line) for all fields in this survey which 
have plate centres with dec$\geq$-30$^{\circ}$}
\end{center}
\end{figure}

\begin{figure}
\begin{center}
\begin{minipage}{6in}
   \epsfig{file=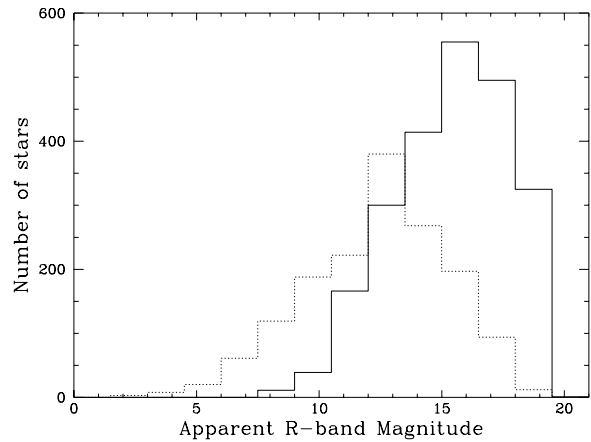,width=0.4\hsize,angle=270}
\end{minipage} 
  \caption{Same as Fig.6 but for fields which have plate centres with dec$\leq$-40$^{\circ}$}
\end{center}
\end{figure}

\section{Object populations and selection for follow up observations}

As noted in Sect. 1$\cdot$1, proper motion surveys can give an insight into both the local stellar neighbourhood 
and halo populations.  In order to have a high proper motion an object must be either nearby or have a high 
space velocity (or both).  High proper motion surveys therefore naturally select nearby and halo stars.  
Analysis of the objects identified by the survey requires differentiation of the populations, and this can be 
achieved by use of reduced proper motion and colour-colour diagrams.  Astrophysically interesting objects can 
then be selected for follow up observations on this basis.  

We recently obtained data for a number of stars selected from the catalogue on the ESO New Technologies 
Telescope at the La Silla observatory, Chile.  The stars were selected as; those with the highest proper 
motions, candidate brown dwarfs, halo subdwarfs.  These data are currently being reduced.

\subsection{Reduced Proper Motion}

The concept of reduced proper motion (H) was first used by Ejnar Hertzsprung (Luyten, 1968) as an alternative 
to absolute magnitude in order to produce `statistical' HR diagrams.  It combines the observable properties 
of proper motion, $\mu$, and apparent magnitude, $m$, via the equation,
\begin{equation}
	H=m+5+5log\mu
\end{equation}
From the definition of parallax it can be seen that,
\begin{equation}
	V_T=4.74{\mu\over\pi}
\end{equation}
where $V_T$ is the tangential velocity measured in kms$^{-1}$, and $\pi$ is the 
parallax in arcseconds, and since,
\begin{equation}
	\pi={1\over d}
\end{equation}
where $d$ is in parsecs, it follows that,
\begin{equation}
	d={V_T\over 4.74\mu}.
\end{equation}
Using this and the relation between apparent and absolute magnitudes,
\begin{equation}
	m=M+5log\left({d\over 10}\right),
\end{equation}
we can substitute into eqn.5 to get,
\begin{equation}
	H=M+5log{V_T}-3.379.
\end{equation}
From this it can be seen that the reduced proper motion, as calculated from the measured properties of proper 
motion and apparent magnitude, is equivalent to a combination of the absolute magnitude and tangential 
velocity, which are intrinsic properties of a star.  Plotting the reduced proper motion against colour thus 
gives a plot which is in essence an H-R diagram modified to include tangential velocity, called a reduced 
proper motion diagram.

One of the advantages of such a plot is that the higher mean tangential velocity of halo subdwarfs, when 
combined with their intrinsic subluminosity makes them far easier to differentiate from the disk population.  
However, the spread in H of objects in a reduced proper motion diagram is far larger than the spread in 
magnitude in a standard H-R diagram because of the random spread of velocities about the mean values for the 
different populations.  It also means that stars can be misidentified due to unusually high or low tangential 
velocities.

Fig. 8 is a plot of the B-band reduced proper motion (H$_B$) against B-I and shows the differentiation 
between the white dwarf, halo and disc populations.  It should be noted that for $\sim$150 faint red stars no 
matches could be found on the B$\rm{_J}$ plates.

\begin{figure*}
\begin{center}
\begin{minipage}{6in}
   \epsfig{file=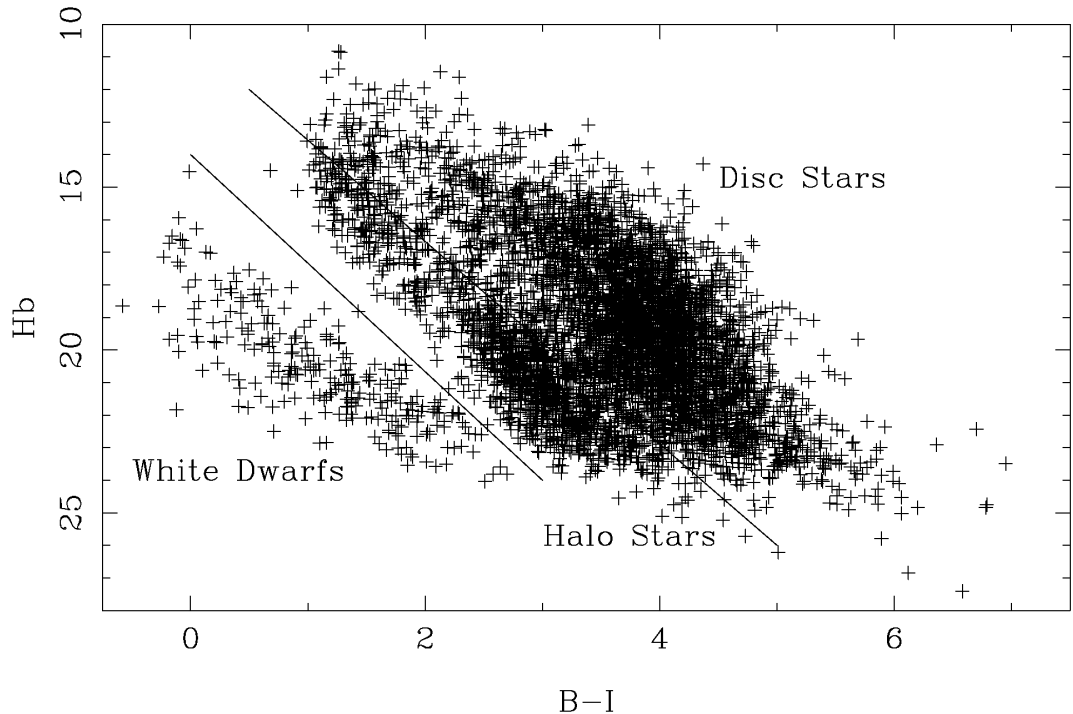,width=0.8\hsize,angle=270}
\end{minipage} 
  \caption{Plot of H$_B$ against B--I for all stars in the survey with B and I magnitudes.  The 
lines indicating the different regions of the plot are approximate and for guidance only (see Sect. 5$\cdot$1).}
\end{center}
\end{figure*}

\subsection{The Colour-Colour Diagram}

The BRI two colour diagram (Fig. 9) utilises the B$\rm{_J}$ and I plates from the UK Schmidt 
survey.  Halo subdwarfs and giant stars occupy the bottom right of this plot, the white dwarfs 
form the tail to the lower left and the disc main sequence runs from above the halo subdwarf 
region to the top right of the plot.  This plot can therefore be used in conjunction with the 
reduced proper motion plots to differentiate between the various populations of objects.

\begin{figure*}
\begin{center}
\begin{minipage}{6in}
   \epsfig{file=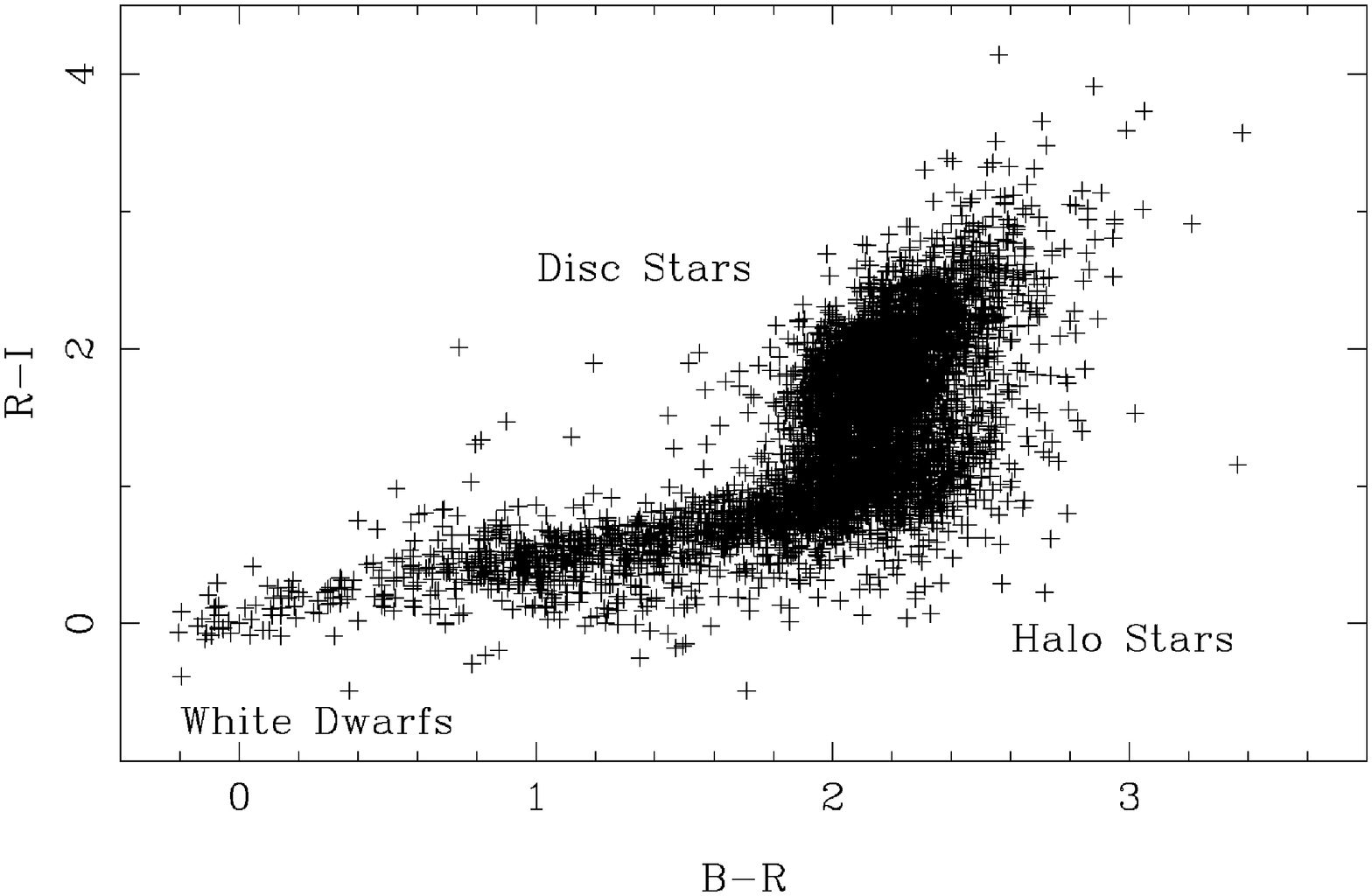,width=0.8\hsize,angle=270}
\end{minipage} 
  \caption{Plot of B--R against R--I for all stars in the survey with B and I magnitudes}
\end{center}
\end{figure*}

\section{Discussion}

Over the past few years there have been several papers which offer a variety of values for the 
completeness of the Luyten catalogues, and discuss its shortcomings (see Sect. 1$\cdot$2).  The values 
offered for the completeness of the Luyten catalogues vary from 60$\%$ for m$_R$$>$13 (Ruiz et al., 
2001,comparison to Calan-ESO survey conducted at low galactic latitude) to 90$\%$ for m$_R$$\le$18 
and $\mu$$\ge$0$\cdot$5$\arcsec$yr$^{-1}$ (Dawson 1986, V/V$\rm{_{max}}$ method applied to northern areas 
of LHS)

Our completeness levels are the first to be calculated from an analysis of the search technique 
itself, and show that for the vast majority of stars in the catalogue (R$\le$19mag and 
0$\cdot$2$\le$$\mu$$\le$2$\cdot$5$\arcsec$yr$^{-1}$) the completeness is greater than 90$\%$.  
Fig. 6 therefore indicates that for the Northern areas ($\delta$$>$-33$^{\circ}$, 
$\mid$b$\mid$$>$10$^{\circ}$) the NLTT is $\sim$85$\%$ complete.  Fig. 7 shows the 
incompleteness of the Southern part of the NLTT for R$>$13mag ($\sim$25$\%$ complete 
for 15$\le$R$\le$16$\cdot$5mag and $\sim$15$\%$ for 16$\cdot$5$\le$R$\le$18mag).

We used the H$_B$, B-I and the B--R, R--I plots to select halo subdwarfs.  This selection yielded 
1189 halo subdwarfs from a total of 6104 stars which have both B and I band magnitudes ($\sim$19.5$\%$).  
Table 6 of Reid (1984) gives four model predictions of the relative numbers of disc, halo, intermediate 
(thick disc) and white dwarf populations which would be discovered by a polar proper-motion survey.  
The models are based on different combinations of the disc and halo luminosity functions and the 
kinematics of the halo.  Our result gives a value which falls between the predictions of models A 
and D and excludes models B amd C.  Model A is based largely on the results of Wielen (1974), whilst 
model D uses the Luyten (1968) disc luminosity function, a `mean globular' halo luminosity function 
and the halo kinematics of Oort (1965).  This analysis is described in more detail in Pokorny, Jones \& 
Hambly, 2002 (in preparation).

\section*{Acknowledgments}
The authors would like to thank all of the staff of the SuperCOSMOS project at the ROE and Mike Read for 
producing Fig. 1.  RSP would like to thank J. M. Porter, P. A. James and A. M. Newsam for helpful discussions 
and suggestions.  RSP acknowledeges support of an LJM University fellowship.


\newpage
\begin{thebibliography}{30} 
\bibitem{b1} Dawson P.C., 1986, ApJ, 311, 984.
\bibitem{b2} Evans D.W., 1992, MNRAS, 255, 521.
\bibitem{b3} Flynn C., Sommer-Larsen J., Fuchs B., Graff D.S., \& Salim S., 2001, MNRAS, 322, 553.
\bibitem{b4} Hambly N. C., Miller L., MacGillivray H. T., Herd J. T., \& Cormack, W. A., 1998, MNRAS, 298, 897.
\bibitem{b5} Hambly N.C., MacGillivray H.T., Read M.A., et al., 2001a, MNRAS, 326, 1279.
\bibitem{b6} Hambly, N.C., Irwin, M.J., \& MacGillivray, H.T., 2001b, MNRAS, 326, 1295.
\bibitem{b7} Hambly, N.C., Davenhall, A.C., Irwin, M.J., \& MacGillivray, H.T., 2001c, 
MNRAS, 326, 1315.
\bibitem{b8} Henry T.J., Ianna P.A., Kirkpatrick J.D., \& Jahreiss H., 1997, AJ, 114, 388.
\bibitem{b9} Lee S-G., 1993, JKAS, 26, 141.
\bibitem{b10} Luyten W.J., 1968, MNRAS, 139, 221.
\bibitem{b11} Luyten W.J., 1979, LHS catalogue, University of Minnesota.
\bibitem{b12} Luyten W.J., 1979, NLTT catalogue, University of Minnesota.
\bibitem{b13} Monet D.G., Fisher M.D., Liebert J., et al., 2000, AJ, 120, 1541.
\bibitem{b14} Morgan D.H., Tritton S.B., Savage A., Hartley M., \&Cannon R.D., 1992, in: Digitised Optical Sky Surveys, eds. H.T. McGillivray \& E.B. Thomson (Kluwer: Dordrecht), 11.
\bibitem{b15} Oort, J.H., 1965, Galactic Structure, ch. 15, University of Chicago Press.
\bibitem{b16} Reid, I.N., 1984, MNRAS, 206, 1.
\bibitem{b17} Ruiz M.T., Leggett S.K., \& Allard F., 1997, ApJ, 491, 107.
\bibitem{b18} Ruiz M.T., Wischnjewsky, M., Rojo P.M., \& Gonzalez L.E., 2001, ApJS, 133, 119.
\bibitem{b19} Schmidt M., 1968, ApJ, 151, 393.
\bibitem{b20} Schmidt M., 1975, ApJ, 202, 22.
\bibitem{b21} Scholz R.-D., Irwin W., Ibata R., Jahreiss H., \& Malkov O.Yu., 2000, A\&A, 353, 958.
\bibitem{b22} Wielen R., 1974, Highlights of Astronomy, 3, 395.
\bibitem{b23} Williams D. B., 2000, JAAVSO, 28, 12.

\end{thebibliography}
\end{document}